\RequirePackage{fix-cm}
\documentclass[twocolumn,epjc3]{svjour3}  
\usepackage{newtxtext,newtxmath} 
\RequirePackage{graphicx}
\RequirePackage{subfigure}
\RequirePackage{rotating}
\RequirePackage{pdflscape}
\RequirePackage{color}
\usepackage{mathtools}
\usepackage{extarrows}
\usepackage{bm}
\RequirePackage{latexsym}
\RequirePackage{xspace}
\RequirePackage[numbers,sort&compress]{natbib}
\RequirePackage[colorlinks,citecolor=blue,urlcolor=blue,linkcolor=blue]{hyperref}
\usepackage{amsmath,amsfonts,bm}
\usepackage{wasysym}
\usepackage{color}
\usepackage{lineno}
\usepackage[usenames,dvipsnames,svgnames,table]{xcolor}	
\let\oldequation\equation
\let\oldendequation\endequation

\renewenvironment{equation}
  {\linenomathNonumbers\oldequation}
  {\oldendequation\endlinenomath}

\let\oldalign\align
\let\oldendalign\endalign

\renewenvironment{align}
  {\linenomathNonumbers\oldalign}
  {\oldendalign\endlinenomath}

\usepackage{acronym}
\usepackage{xcolor}
\usepackage{braket}

\journalname{Eur. Phys. J. C}

\begin{document}

\title{Precise $^{113}$Cd $\beta$ decay spectral shape measurement and interpretation in terms of possible $g_A$ quenching}
\author{
I.~Bandac\thanksref{LSC}\and
L.~Berg\'e\thanksref{IJCLab}\and
J.M.~Calvo-Mozota\thanksref{LSC,ESICTV,ESIT}\and
P.~Carniti\thanksref{INFN-Milano}\and
M.~Chapellier\thanksref{IJCLab}\and
F.~A.~Danevich\thanksref{KINR,INFN-Roma}\and
T.~Dixon\thanksref{IJCLab,e1}\and
L.~Dumoulin\thanksref{IJCLab}\and
F.~Ferri\thanksref{CEA-IRFU}\and
A.~Giuliani\thanksref{IJCLab}\and
C.~Gotti\thanksref{INFN-Milano}\and
Ph.~Gras\thanksref{CEA-IRFU}\and
D.~L.~Helis\thanksref{CEA-IRFU,e2}\and
L.~Imbert\thanksref{e3,IJCLab}\and
H.~Khalife\thanksref{CEA-IRFU}\and 
V.~V.~Kobychev\thanksref{KINR}\and
J.~Kostensalo\thanksref{NRIF}\and
P.~Loaiza\thanksref{IJCLab}\and
P.~de~Marcillac\thanksref{IJCLab}\and
S.~Marnieros\thanksref{IJCLab}\and
C.A.~Marrache-Kikuchi\thanksref{IJCLab}\and
M.~Martinez\thanksref{CAFAE,ARAID} \and
C.~Nones\thanksref{CEA-IRFU}\and
E.~Olivieri\thanksref{IJCLab}\and
A.~Ortiz~de~Sol\'orzano\thanksref{CAFAE}\and
G.~Pessina\thanksref{INFN-Milano}\and
D.~V.~Poda\thanksref{IJCLab}\and
J.A.~Scarpaci\thanksref{IJCLab}\and
J.~Suhonen\thanksref{Jyvaskyla, ICATRP-Romania}\and
V.~I.~Tretyak\thanksref{KINR, LNGS}\and 
M.~Zarytskyy\thanksref{KINR}\and
A.~Zolotarova\thanksref{CEA-IRFU}
}

\thankstext{e1}{Currently at University College London, London, UK}
\thankstext{e2}{Currently at Laboratori Nazionali del Gran Sasso, Assergi, Italy}
\thankstext{e3}{e-mail: leonard.imbert@ijclab.in2p3.fr}

\institute{Laboratorio Subterr\'aneo de Canfranc, 22880 Canfranc-Estaci\'on, Spain \label{LSC}\and
Université Paris-Saclay, CNRS/IN2P3, IJCLab, 91405 Orsay, France \label{IJCLab} \and
Escuela Superior de Ingenier\'ia, Ciencia y Tecnolog\'ia, Universidad Internacional de Valencia – VIU, 46002 Valencia, Spain \label{ESICTV}\and
Escuela Superior de Ingenier\'ia y Tecnolog\'ia, Universidad Internacional de La Rioja, 26006 Logro\~no, Spain \label{ESIT}\and
INFN, Sezione di Milano-Bicocca, I-20126 Milano, Italy \label{INFN-Milano} \and 
Institute for Nuclear Research of NASU, 03028 Kyiv, Ukraine \label{KINR} \and 
INFN, Sezione di Roma, P.le Aldo Moro 2, I-00185, Rome, Italy \label{INFN-Roma} \and
IRFU, CEA, Universit\'{e} Paris-Saclay, F-91191 Gif-sur-Yvette, France  \label{CEA-IRFU} \and 
Natural Resources Institute Finland, Yliopistokatu 6B, FI-80100 Joensuu, Finland \label{NRIF} \and
Centro de Astropart\'iculas y F\'isica de Altas Energ\'ias, Universidad de Zaragoza, Zaragoza 50009, Spain \label{CAFAE}\and
ARAID Fundaci\'on Agencia Aragonesa para la Investigaci\'on y el Desarrollo, 50018 Zaragoza, Spain \label{ARAID}\and
Department of Physics, University of Jyv\"askyl\"a, P.O. Box 35, FI-40014, Jyv\"askyl\"a, Finland \label{Jyvaskyla}  \and
International Centre for Advanced Training and Research in Physics, P.O. Box MG12, 077125 Bucharest-M{\u a}gurele, Romania \label{ICATRP-Romania} \and
INFN, Laboratori Nazionali del Gran Sasso, I-67100 Assergi (AQ), Italy \label{LNGS}
}




\date{Received: date / Accepted: date}

\maketitle

\begin{abstract}

Highly forbidden $\beta$ decays  provide a sensitive test to nuclear models in a regime in which the decay goes through high spin-multipole states, similar to the neutrinoless double-$\beta$ decay process. There are only 3 nuclei ($^{50}$V, $^{113}$Cd, $^{115}$In) which undergo a $4^{\rm th}$ forbidden non-unique $\beta$ decay. 
In this work, we compare the experimental $^{113}$Cd spectrum to theoretical spectral shapes in the framework of the spectrum-shape method. We measured with high precision, with the lowest energy threshold and the best energy resolution ever, the $\beta$ spectrum of $^{113}$Cd embedded in a 0.43 kg CdWO$_4$ crystal, operated over 26 days as a bolometer at low temperature in the Canfranc underground laboratory (Spain). We performed a Bayesian fit of the experimental data to three nuclear models (IBFM-2, MQPM and NSM) allowing the reconstruction of the spectral shape as well as the half-life. The fit has two free parameters, one of which is the effective weak axial-vector coupling constant, $g_A^{\text{eff}}$, which resulted in $g_A^{\text{eff}}$ between 1.0 and 1.2,  compatible with a possible quenching. Based on the fit, we measured the half-life of the $^{113}$Cd $\beta$ decay including systematic uncertainties as  $7.73^{+0.60}_{-0.57} \times 10^{15}$ yr, in agreement with the previous experiments. These results represent a significant step towards a better understanding of low-energy nuclear processes.

\end{abstract}

\section{Introduction}
\label{intro}

Theoretical calculations overpredict the decay rate of some $\beta$ and two-neutrino double-$\beta$ ($2\nu\beta\beta$) processes over a wide set of observations  \cite{Suhonen:2017,Suhonen:2013laa} for which the calculated nuclear matrix elements (NMEs) are too large to reproduce the experimental rates. This reveals a deficiency of the nuclear models, which could be related to nuclear medium effects like the lack of two-body currents, three-nucleon forces, and valence and configuration-space effects. One possibility to solve the discrepancy would be to renormalize the axial-vector coupling strength to a lower value with respect to that of a free nucleon $g_A = 1.276$ \cite{Mund:2012fq,Markisch:2018ndu}.

Neutrinoless double-$\beta$ decay ($0\nu\beta\beta$) is a hypothetical nuclear process, which, if observed, would demonstrate that the neutrino is the only fermion to be a Majorana particle and would result in the violation of the lepton number conservation by two units \cite{Agostini:2022zub, DellOro:2016tmg, gomez2023search}. At present, the double-$\beta$ community is building next-generation experiments capable to reach sensitivities of ${\sim}10^{27}$ yr on the $0\nu\beta\beta$ half-life. The correlation between single-$\beta$ or $2\nu\beta\beta$ NMEs and $0\nu\beta\beta$ decay NMEs is not straightforward. 
However, the authors in \cite{Jokiniemi:2022ayc} observed good linear correlations between $2\nu\beta\beta$ and $0\nu\beta\beta$ decay NMEs for shell-model and proton-neutron quasi-particle random-phase approximation (pnQRPA) calculations for several tens of decays.
The effect of a renormalization of the NMEs or more generally, the uncertainty on the NMEs, often folded with the value of $g_A$, would strongly impact the predicted $0\nu\beta\beta$ decay rates. 

The rate for the $0\nu\beta\beta$ decay, in case of a light Majorana neutrino exchange, can be written as:
\begin{equation}
    \left(T^{0\nu}_{1/2} \right)^{-1} =(g_A^{0\nu})^4\cdot G_{0\nu}\cdot |M_{0\nu}|^2\cdot \langle m_{\beta\beta}\rangle^2/m_e^2,
\end{equation}
where $g_A^{0\nu}$ is the axial-vector coupling strength for the $0\nu\beta\beta$ decay, $G_{0\nu}$ is the phase space factor, $M_{0\nu}$ is the NME, $ \langle m_{\beta\beta} \rangle$ is the effective Majorana neutrino mass and $m_e$ is the electron mass.
Improving the accuracy of $0\nu\beta\beta$ NMEs depends on tuning and validating the nuclear models using experimental data. Measurements of the $2\nu\beta\beta$ decay \cite{KamLAND-Zen:2019imh, CUPID-Mo:2023lru, GERDA:2023wbr}, a process which has the same initial and final state as  $0\nu\beta\beta$ decay,  can provide information on the structure of the parent and daughter nucleus. Complementary measurements are obtained from muon capture \cite{Kortelainen:2002bz, Bajpai:2024bkf}, which involves similar momentum transfer as $0\nu\beta\beta$ decay.
Highly forbidden $\beta$ decays can test  nuclear models in the regime of large angular momentum differences between the nuclear states. This is similar to the $0\nu\beta\beta$ which has a large momentum transfer and proceeds via higher-multipolarity  intermediate nuclear states.
Thus, it is necessary to understand the standard processes to make confident predictions and extract information on beyond standard model physics from the $0\nu\beta\beta$ decay.\\

In this work, we report a measurement of the $\beta$ spectrum shape of $^{113}$Cd ($Q_{\beta} = 323.84(27)$ keV \cite{Wang:2021xhn}) using the cryogenic calorimeter technique which offers excellent energy resolution, low energy threshold and high detector efficiency. 
Massive cryogenic calorimeters can achieve energy thresholds of $\sim$1--10 keV and allow to discriminate $\alpha$ from $\gamma/\beta$ particles when they feature scintillation capabilities \cite{Poda:2021hsv} with a double readout.
We studied $^{113}$Cd embedded in a cadmium tungstate (CdWO$_4$) crystal with a natural abundance measured precisely as ($12.22 \pm 0.02$)\% \cite{Belli:2007zza}. 
The same crystal was previously instrumented with photomultiplier tubes \cite{Belli:2007zza} and employed in a scintillation low-background experiment in which the most precise value of the $^{113}$Cd half-life $T_{1/2} = (8.04 \pm 0.05) \times 10^{15}$ yr and the $\beta$-spectrum shape were measured \cite{Belli:2007zza}. The sensitivity of highly forbidden non-unique $\beta$ decays to test the nuclear models was shown in \cite{Haaranen:2017, Kumar:2020, Mustonen:2006qn}. 
The spectrum-shape method, SSM, was proposed in \cite{Haaranen:2016rzs} to measure an effective value $g_A^{\text{eff}}$. The method was updated by including the half-life in the model and, so far, it has been applied by the COBRA experiment on the $^{113}$Cd decay  \cite{Kostensalo:2020gha}, and by \cite{Pagnanini:2024qmi} and \cite{Denys_In115} on the $^{115}$In decay. In our work, we employ an analysis relying on a Bayesian fit to extract the value of the $g_A^{\text{eff}}$, based on the same theoretical framework as in \cite{Kostensalo:2020gha}, including an extensive study of the systematic uncertainties. Compared to previous studies on $^{113}$Cd $\beta$ decay, the bolometric technique allows to obtain a lower energy threshold and a better energy resolution.
We describe in Section \ref{sec:theory}  the theoretical framework and in Section \ref{sec:measurement} we present the data analysis performed to extract the energy spectrum of the CdWO$_4$ crystal. In Section \ref{sec:BM} we report the background model fit that results in a global background spectrum, used in the Bayesian fit to extract the value of the $g_A^{\text{eff}}$, detailed in Section \ref{sec:Analyis} together with the results.
\section{Theoretical framework}
\label{sec:theory} 

In general, for the $\beta$ decay, one can write the differential probability of the electron energy as \cite{Haaranen:2017, Kumar:2020, Mustonen:2006qn}:
\begin{align}
\label{eq:general_beta_decay}
    P(w_e) dw_e = \frac{ \left( G_F \cos \theta \right) ^2}{\left( \hbar c \right)^6} \frac{1}{2 \pi^3 \hbar} C(w_e) p_e c w_e (w_0 - w_e)^2 \\ \nonumber F_0(Z,w_e) dw_e,
\end{align}
where $G_F$ is the Fermi coupling constant, $\theta$ is the Cabibbo angle, $C(w_e)$ is a nuclear shape factor, $w_0$ is the end-point energy of the process, $F_0(Z,w_e)$ is the Fermi function with $Z$  the number of protons in the daughter nucleus, $p_e$ and $w_e$ are respectively the momentum and the energy of the electron. In pure Gamow-Teller transition, the shape factor $C(w_e)$ is relatively simple \cite{Kumar:2020}. On the contrary, in the case of forbidden non-unique $\beta$ decay, the nuclear shape factor can be written as \cite{Haaranen:2017, Kumar:2020, Mustonen:2006qn}:
\begin{align}
    \label{eq:gA_general_shape}
    C(w_e) = \sum_{k_\nu,k_e,K} \lambda_{k_e} \Big[ M_K^2(k_e,k_\nu) + m_K^2(k_e,k_\nu) \\ \nonumber  - \frac{2 \gamma_{k_e}}{k_e w_e} M_K(k_e,k_\nu)  m_K(k_e,k_\nu) \Big],
\end{align}
where $k_e$ and $k_\nu$ are positive integers related to the partial-wave expansion of the lepton wave functions, and $K$ is the order of forbiddeness, $\lambda_{k_e}$ is the Coulomb function, $\gamma_{k_e} = \left( k_e^2 - (\alpha Z)^2 \right)^{1/2}$, with $\alpha$ = 1/137, the fine structure constant, and $M_K(k_e,k_\nu)$ and $m_K(k_e,k_\nu)$ contain the NMEs and phase space factors. These last two parameters are evaluated within the impulse approximation, which considers that the decay occurs in the neutron independently from the other nucleons. They can be understood as an intricate mixture of phase space factors and NMEs. Schematically, one can write \cite{Haaranen:2017, Kumar:2020, Mustonen:2006qn}:
\begin{align}
    \label{eq:gA_MK}
    M_K(k_e,k_\nu) \propto (w_0 - w_e)^{k_{\nu}-1} \Big[ g_V f^{(0)}_{K,K-1,1}  \ ^{V} \mathcal{M}_{K,K-1,1}^{(0)}  \\ \nonumber  - g_V f^{(0)}_{K,K,0} \ ^{V} \mathcal{M}_{K,K,0}^{(0)} \\ \nonumber +  g_A f^{(0)}_{K,K,1} \ ^{A} \mathcal{M}_{K,K,1}^{(0)}  \\ \nonumber  + ... \Big],
\end{align}
and:
\begin{align}
     \label{eq:gA_mK}
     m_K(k_e,k_\nu) \propto (w_0 - w_e)^{k_{\nu}-1} \Big[ g_V h^{(0)}_{K,K,0} \ ^{V} \mathcal{M}_{K,K,0}^{(0)}  \\ \nonumber  - g_V h^{(0)}_{K,K-1,1} \ ^{V} \mathcal{M}_{K,K-1,1}^{(0)} \\ \nonumber + g_A h^{(0)}_{K,K,1} \ ^{A} \mathcal{M}_{K,K,1}^{(0)}  \\ \nonumber  + ... \Big],
\end{align}
where $f$ and $h$ are the corresponding phase space factors for each of the NMEs $\mathcal{M}$, and $g_A$ and $g_V$ are the weak axial-vector and vector coupling constants, respectively.
The parameter $\mathcal{M}_{K,L,S}^{(0)}$ corresponds to the NME:
\begin{align}
    ^{V/A} \mathcal{M}_{K,L,S}^{(0)} \propto \sum_{p,n} \ ^{V/A}m_{K,L,S} \bra{p} | \mathcal{O}^{(0)}_{K,L,S} | \ket{n} \\ \nonumber
    \bra{\psi_f} | [c_p^\dagger \tilde{c_n}]_K | \ket{\psi_i}  ,
\end{align}
where the sum runs over the neutrons $n$ and the protons $p$. The first part under the sum is the single-particle matrix ele\-ment characterizing the transition operator $\mathcal{O}^{(0)}_{K,L,S}$ and the second part is the one-body transitions density characterizing the nuclear structure through the initial ($\psi_i$) and final ($\psi_f$) nuclear wave functions. The matrix elements are related to the unitless form factors $F^{(0)}_{KLS}$ by the equations:
\begin{align}
    R^L \ ^V F^{(0)}_{KLS} = (-1)^{K-L} g_V \ ^V M^{(0)}_{KLS}, \\
    R^L \ ^A F^{(0)}_{KLS} = (-1)^{K-L+1} g_A \ ^A M^{(0)}_{KLS}, 
\end{align}
where $R$ is the nuclear radius. This gives the matrix elements the unit fm$^L$.
Finally, the nuclear shape factor $C(w_e)$ depends on the $g_A$, the $g_V$, and mixed terms of $g_A$ $g_V$, due to the power of two on the terms $M_K$ and $m_K$, in Eq. \ref{eq:gA_general_shape}.

This reveals that the spectral shape of highly forbidden non-unique $\beta$ decay depends on the value of the $g_A$. The disintegration of $^{113}$Cd corresponds to a four-fold forbidden non-unique $\beta$ decay and its $\beta$ spectral shape is sensitive to the ratio $g_{A}/g_V$. Assuming that the vector current in the V-A theory of weak interactions is conserved, with $g_V = 1$, the $\beta$ spectral shape is thus sensitive to the value of axial coupling $g_A$ and it can be probed through the spectrum-shape method \cite{Haaranen:2016rzs}. In this method, the $g_A$ is not considered as a fundamental constant but rather an effective factor, specific for each nuclear model, that can relate to a lack of correlations or not large enough considered valence spaces in the models. Thus, we introduce the effective axial-vector coupling constant $g_A^{\text{eff}}$, which is the parameter one can measure experimentally within this method. In a simplified view, one can write: 
\begin{equation}
    C(w_e) = g_V^2 C_V(w_e) + \left(g_A^{\text{eff}}\right)^2 C_A(w_e) + g_V g_A^{\text{eff}} C_{VA}(w_e),
\end{equation}
where $g_V = 1$ in the conserved vector current hypothesis, and $C_V(w_e)$, $C_A(w_e)$, $C_{VA}(w_e)$ are respectively the shape factor of the vector part, the axial-vector part and the mixed vector-axial vector part. In the expressions \ref{eq:gA_MK} and \ref{eq:gA_mK}, the dominating contribution comes from $\mathcal{M}_{K,K,0}^{(0)}$ and can be calculated with a good precision. However, other terms can still play a non-negligible role, like the NME $\mathcal{M}_{K,K-1,1}^{(0)}$ which is difficult to calculate. In particular, within the NSM (nuclear shell model) and IBFM-2 (microscopic interacting boson-fermion model) calculations, like in \cite{Kostensalo:2020gha}, this contribution is found to be zero due to the small size of the single-particle space. The MQPM (microscopic quasiparticle-phonon model) can handle a larger single-particle valence space, giving a value of $\sim$0.4 for this contribution \cite{Kostensalo:2020gha}. Under the conserved vector current hypothesis, it is possible to evaluate these NMEs. In particular for the $^{113}$Cd decay we have, $\mathcal{M}_{4,3,1}^{(0)}$(IBFM-2) = 3.7~fm$^3$, $\mathcal{M}_{4,3,1}^{(0)}$(MQPM) = 9.3~fm$^3$, and $\mathcal{M}_{4,3,1}^{(0)}$(NSM) = 8.4~fm$^3$ \cite{Kostensalo:2023xzu}. Nevertheless, fixing this value does not allow to reproduce at the same time the spectral shape as well as the half-life of the decay, as it was observed in \cite{Cobra:2020} and \cite{Denys_In115}. 
Recently, the SSM was improved by considering the NME $\mathcal{M}_{K,K-1,1}^{(0)}$, also called a small relativistic NME, s-NME, as a free parameter. This allows to match the half-life as well as the spectral shape to those of experimental data. The  COBRA experiment recently implemented this enhanced  SSM \cite{Kostensalo:2020gha}. 
In our work, we consider the enhanced SSM, and we used two free parameters, $g_A^{\text{eff}}$ and the s-NME, to fit the spectral shape and the half-life at the same time.

\section{Experimental data}
\label{sec:measurement}




The CdWO$_4$ crystal (CWO) was operated in the CROSS cryogenic facility in the underground laboratory of Canfranc (LSC). The CWO has a cylindrical shape with a diameter of $40$ mm and a length of $43$ mm, for a total mass of $433.61$~g. It was operated in a dual readout mode, measuring the heat and the scintillation light. The CWO was instrumented with a Neutron Transmutation Doped Ge thermistor (NTD) to read the heat signal, producing pulses of $\sim1$ s length, and a heater periodically injecting a given energy to correct temperature fluctuations. The CWO crystal was facing a light detector (LD) consisting of a Ge wafer also operated as a cryogenic calorimeter. The LD exploits the Neganov--Trofimov--Luke effect, which enhances the signal-to-noise ratio \cite{Neganov:1985khw,Luke:1988}. The LD is also instrumented with an NTD, which reads a phonon signal induced by the scintillation signal from CWO. The part of the LD facing the crystal was coated with an anti-reflecting SiO layer to increase the light collection.  The crystal was held by PTFE pieces to ensure the coupling to the thermal bath, and surrounded by a copper frame. The lateral and the bottom parts of the cryostat are shielded from environmental $\gamma$'s by $25$ cm of lead, and the top part by $13$ cm of lead. The  CWO crystal was installed, together with five other crystals of different types (see in \cite{Bandac:2023pkk}), in the CROSS dry cryostat. Studies of the performance of Li$_2$MoO$_4$ (LMO) crystals with natural/enriched/depleted $^{100}$Mo content and $^{116}$Cd-enriched CWO present in the same set-up were reported in \cite{Bandac:2023pkk, CROSS:2023xdt, CUPID:2020pra, Helis:2020ydg}. In this analysis, we used the  $^{100}$Mo-depleted LMO crystal for the background modelling as presented in Section \ref{sec:BM}. More details on the set-up and the data taking can be found in \cite{Bandac:2023pkk}.


Data were acquired between February and April 2020 at a temperature of 12 mK. We registered continuous data streams measuring the output voltage of the NTD with a sampling rate of $2$ kHz. The data processing was done offline using a MATLAB-based analysis tool \cite{Mancuso:2016wab}. The program triggers and filters the data using the optimum filter that maximises the signal-to-noise ratio using an average signal and noise \cite{Gatti:1986cw}. In the studied bolometer, the pulse shapes are the same for localized and non-localized events. We selected events in the energy range of [750, 3550] keV for the CWO with a window size of 1.5 s, which corresponds to events in the $\gamma$/$\beta$ band having a high signal-to-noise ratio $O(1000)$. We averaged 50 signal-like events to obtain the ``template pulse''.
For the LD it was constructed by averaging 100 pulses in the range that corresponds to [330,~1600] keV in the CWO with a time window of $0.1$ s. The noise power spectrum was built by using 10000 baseline samples without signal events. The triggering of the CWO was done requiring that the pulse amplitude is above 5$\sigma$ of the baseline noise and that Pearson's linear correlation to the mean pulse is $>0.3$. This value corresponds to a very conservative cut, chosen to keep as much as possible the signal events at the level of the processing. 
For each triggered event, we evaluate the signal amplitude that relates to the deposited energy and calculate several pulse-shape parameters. To discriminate the $\alpha$ particles, the coincidence between the CWO and the LD was done using the trigger time of the CWO. We finally obtained $634$ hours of data for the analysis.


The processed data were corrected for possible thermal instabilities during the data taking, referred to as {\it stabilization}, using the heater that produces pulses of the same energy. 
This stabilization is divided into different periods to improve as much as possible the energy resolution. More details on the general process of stabilization in bolometers can be found in \cite{Alessandrello:1998bf}. Another correction was performed by combining the scintillation light measured by the LD with the heat signal to obtain the energy spectrum.


The data were then calibrated using the natural $\gamma$ radio-activity peaks observed in the background, $352$ keV ($^{214}$Pb), $609$ keV, $1120$ keV and $1764$ keV ($^{214}$Bi). These peaks were fitted with a Gaussian function plus a flat background and a linear term:
\begin{equation}
    f(E) = p_1 + p_2 \cdot E + p_3 \cdot e^{-\frac{(E-\mu)^2}{2\sigma^2}}.
    \label{eq:fonction_fit_pic}
\end{equation} 
The peak locations were then fitted to the literature value with a second-order polynomial function with zero intercepts. 

We applied a first basic cut to remove events with a large fluctuation on the baseline. We applied a second cut related to the pulse shape of the signal using an effective $\chi^2$ parameter, which we defined as
\begin{equation}
    \chi^2_{\text{eff}} = \frac{\sum_{i=1}^N \left(J_i - M_i \right)^2}{N},
\end{equation}
where the sum runs over the number of points $N$ (that corresponds to the time window multiplied by the frequency sampling), $J_i$ is the filtered signal value at a given point in time, and $M_i$ is the corresponding value of the average signal. 
We apply an energy-dependent cut to remove spurious peaks and two coincident events in the same waveform that sum together (pile-up).

\begin{figure}[h!]
\centering
\includegraphics[width=0.45\textwidth]{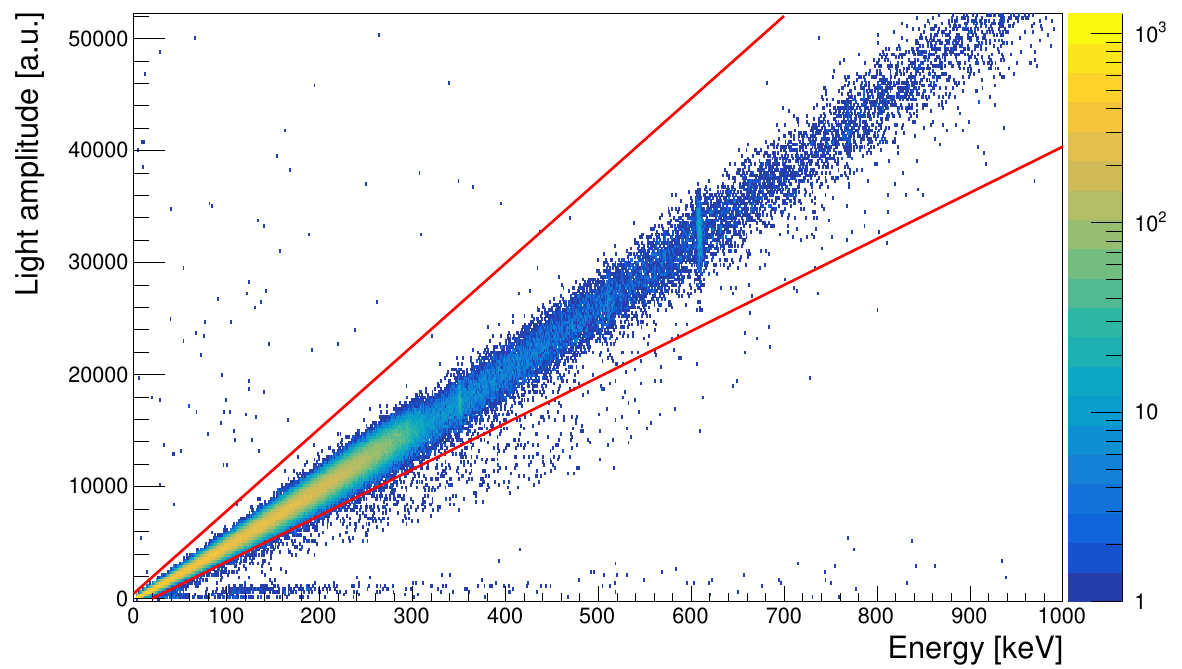}
\caption{2D histogram of experimental data showing the light amplitude as function of the energy. The red lines indicate the selection cut used to remove alpha, nuclear recoils, pile-up and heat-only events.}
\label{fig:light_cut}
\end{figure}

A last cut was applied using the LD that permits to remove $\alpha$ particles and nuclear recoils that have a lower light amplitude than $\beta/\gamma$ particles and heat-only events but also removes some remaining pile-up events as shown in the Fig. \ref{fig:light_cut}. Such events, which cannot be identified as a pile-up by the heat channel due to the time resolution, produce two pulses in the LD that have a better time resolution and are reconstructed with a lower light yield.

\begin{figure}[h!]
\centering
\includegraphics[width=0.45\textwidth]{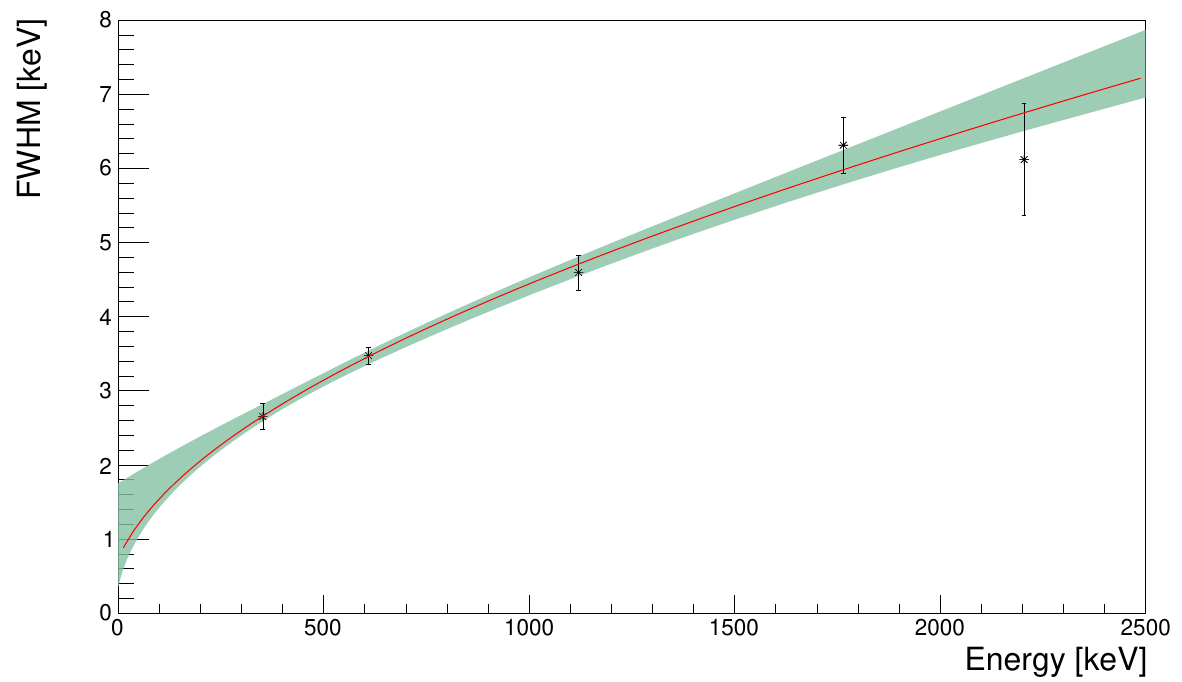}
\caption{Energy resolution (FWHM) as function of the energy. The points are fitted with Eq. \ref{eq:energy_resolution}. The green band shows the $\pm 1 \sigma$ uncertainty of the fit.}
\label{fig:energy_resolution}
\end{figure}

We evaluated the energy resolution of each natural radioactivity $\gamma$ peak observed in the data, by fitting with the same function as in the calibration (Eq. \ref{eq:fonction_fit_pic}). We fitted the obtained energy resolutions with a phenomenological function of the form:
\begin{equation}
    \sigma (E) = \sqrt{\sigma_0^2+p_1E+p_2E^2},
\label{eq:energy_resolution}
\end{equation}
where $\sigma_0$ is related to the baseline noise, and $p_1$ and $p_2$ refer to energy-dependent effects. Figure \ref{fig:energy_resolution} shows the resolution function and the $\pm$1$\sigma$ uncertainty. 

\begin{figure}[h!]
\centering
\includegraphics[width=0.45\textwidth]{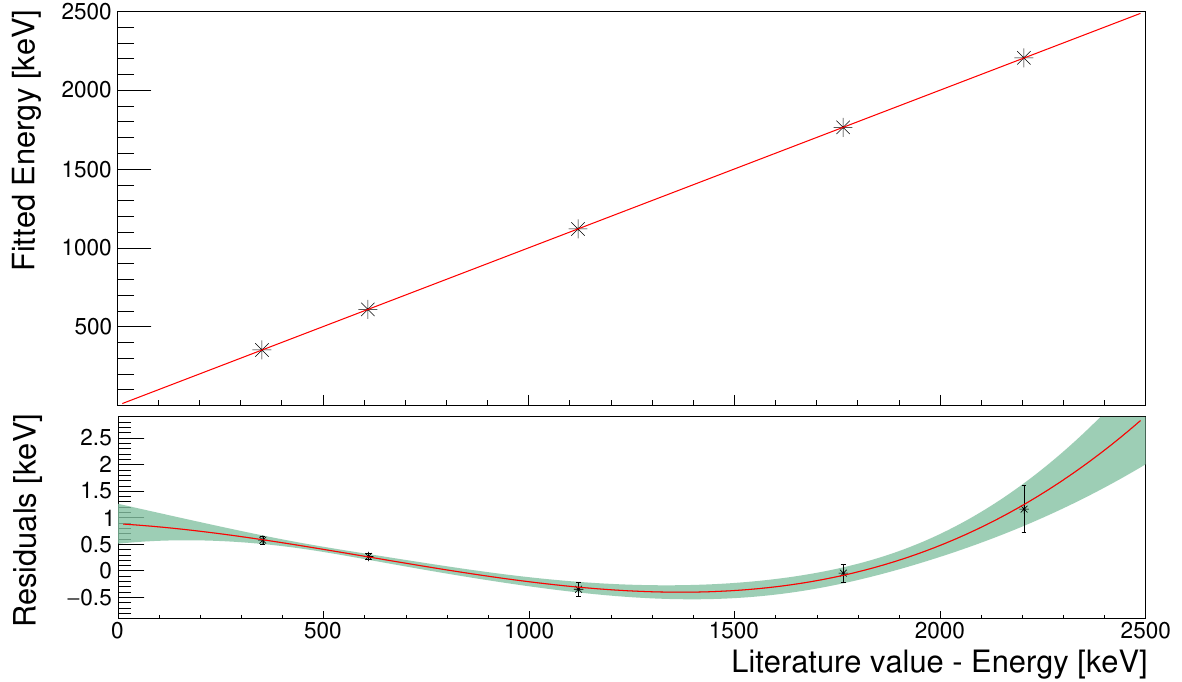}
\caption{Energy bias of the data. The top part shows the fitted energy as function of the literature value for observed $\gamma$ peaks. The bottom part shows the residuals fitted with a third-order polynomial, with the green band indicating the $\pm 1 \sigma$ uncertainty.}
\label{fig:energy_bias}
\end{figure}

We measured the energy bias by comparing the fitted $\gamma$ peaks to the literature value as shown in Fig. \ref{fig:energy_bias}. The bottom panel shows the residuals $\mu - \mu_{\text{lit}}$; we observe a maximum energy bias of $1$ keV in the energy region $<330$ keV, for an extrapolation with a third-order polynomial fit.
The small energy bias is then corrected by subtracting the residual value from each event energy. 

\begin{figure}[h!]
\centering
\includegraphics[width=0.45\textwidth]{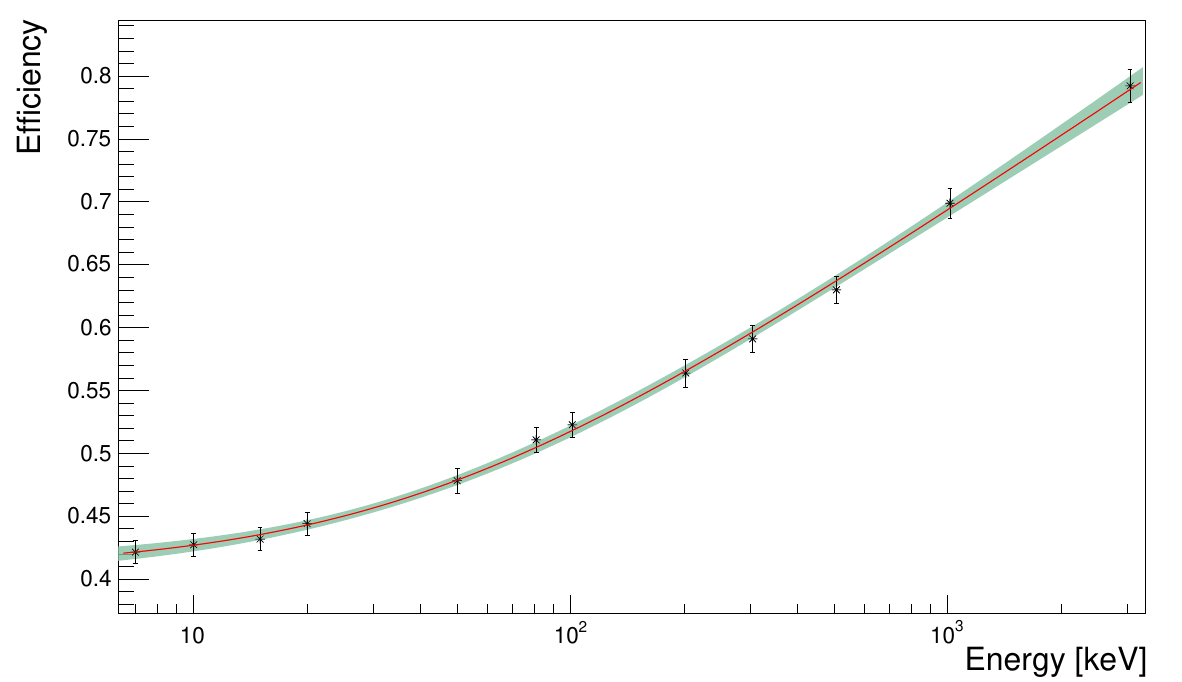}
\caption{Events selection efficiency as function of the energy. The points are obtained from the pulse injection method and are fitted with a phenomenological function given in Eq. \ref{eq:efficiency}. The green band shows the $\pm 1 \sigma$ uncertainty.}
\label{fig:efficiency}
\end{figure}

The event selection efficiency was obtained by the pulse injection method. We constructed an average signal in the same way as described above. We injected this average signal, normalized to a given energy, into the data stream offline. We then applied the same analysis processes as the physics data, i.e. the same processing, calibration and selection cuts. We then count how many events pass all the analysis process to evaluate the efficiency. We injected 12 different amplitudes in the energy range of [7,~3000] keV. We injected 5000 signals for each amplitude to minimize the statistical uncertainty to a level $<2.5\%$. The resulting events selection efficiency as function of the energy is shown in Fig. \ref{fig:efficiency}. The data points are fitted with a phenomenological function defined as 
\begin{equation}
    \epsilon (E) = p_0 \ln{(E+p_1)} + p_2.
\label{eq:efficiency}
\end{equation}
We compared the rise time and decay time distributions of the injected signals and physics data and observed no deviation. In particular, we checked that the injected signals follow the same distribution in the $\chi^2_{\text{eff}}$ variable for the various energies. We did not include the efficiency loss due to the LD cut, which has only a very small effect (it rejects $<1$\% of events).

The pulse injection method was used to determine the analysis threshold, which was defined as the energy for which the tail of the $\chi^2_{\text{eff}}$ distribution of heat-only events in the data is $<$ 1\% from the total data. The analysis threshold is then fixed at $15$ keV.



\section{Background model}
\label{sec:BM}
To extract the spectral shape of the $^{113}$Cd $\beta$ decay, we need to evaluate the background in the energy region $<330$ keV. We performed Monte Carlo (MC) simulations based on GEANT4, version 10.05 \cite{Geant4}. We implemented a detailed geometry of the set-up, including the CWO and the five other crystals, the copper holders, the cryostat screens and the lead shields. We generated the decays of the $^{238}$U and $^{232}$Th  chains using the Decay0 event generator \cite{Ponkratenko:2000um} by considering only the $\beta/\gamma$ emitters. We used the G4EmStandardPhysics model \cite{GEANT42020physics}, and set the production cut lengths to 1 mm. Using the Livermore models and modifying the production cut parameter to lower values did not strongly impact the resulting spectra.

We considered a simple model, where we generated the decays in the CWO crystal, the copper holders, the 10 mK plate (located at the top of the detectors), the gap between the outermost cryostat screen, and the external lead shield, and external environmental $\gamma$'s. The muon flux in the LSC is reduced to (5.26 $\pm$ 0.21) $\times$ 10$^{-3}$ muons m$^{-2}$s$^{-1}$ \cite{Trzaska:2019kuk}. Based on GEANT4 simulations, we evaluated the number of muon-induced events in the energy region below 300 keV to 1 event per hour in the crystal. Thus, this contribution is negligible compared to the activity of the $^{113}$Cd $\beta$ decay in the crystal ($\sim$1000 events per hour) and is not included in the model. In the closer sources, i.e. the CWO crystal, the holders and the 10 mK plate, we generated  $^{214}$Pb and  $^{214}$Bi at equilibrium,  $^{212}$Pb, $^{212}$Bi and $^{208}$Tl at equilibrium and $^{40}$K. We also added $^{210}$Pb and  $^{210}$Bi at equilibrium, and $^{90}$Sr--$^{90}$Y, which could be present in the crystal from anthropogenic origin \cite{Oksana_Polishchuk}. We simulated the $^{87}$Rb beta decay that was observed in \cite{Belli:2007zza}, which has a $Q_{\beta}$ of 282 keV. For the background modelling, we fit the background in the energy region above the $^{113}$Cd beta decay, higher than 330 keV, so the contribution of $^{87}$Rb cannot be constrained by the background model. Thus we included this contribution in the $^{113}$Cd spectral shape fit as discussed in Section \ref{sec:Analyis}. 
In the gap between the outermost cryostat screen and the external shield, we considered only $^{214}$Pb--$^{214}$Bi at equilibrium that are originating from $^{222}$Rn present in the air. We considered the decay of $^{210}$Bi in the lead from $^{210}$Pb ($^{210}$Pb itself is not generated due to its low energy $Q_\beta$ and $\gamma$ particles that are not likely to reach the crystal). For the external environmental $\gamma$'s, we considered $^{40}$K,  $^{214}$Bi and $^{208}$Tl,  the decay product of the other elements are not likely to reach the crystal because of the lead shielding. We added a component for possible remaining pile-up events in the CWO data, which we generated by the convolution of the CWO data with itself. 

We performed a fit to the data with the MC simulations using a binned simultaneous maximum likelihood fit with a Markov Chain Monte Carlo (MCMC) approach \cite{MCMC}, developed by the CUORE and CUPID-0 collaborations \cite{Alduino:2017,Azzolini:2019} using the JAGS software \cite{JAGS_1,JAGS_2}. We took advantage of the LMO crystal produced from molybdenum depleted in $^{100}$Mo, which was also installed in the set-up and has a low internal background to further constrain the background model. We performed a simultaneous fit of these two sets of data with our background model. We fit the CWO spectrum from 330 keV to 3200 keV, with a variable binning with a minimum bin size of 10 keV and a minimum of 30 counts in each bin. The fit of the LMO is done in the range 25--3200 keV and permits to constrain the background in the energy region $<330$ keV. The MC simulations include the event selection efficiency and the energy resolution measured for both the CWO and the LMO crystals. We consider uniform priors on the activity for all the components of the fit.

\begin{figure*}[h!]
\centering
\includegraphics[width=0.9\textwidth]{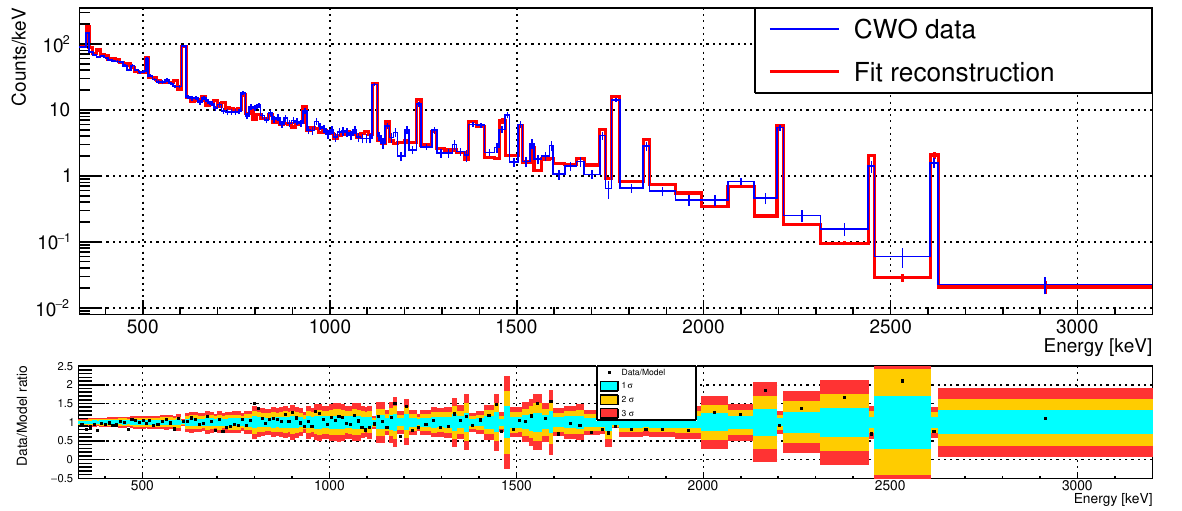}
\includegraphics[width=0.9\textwidth]{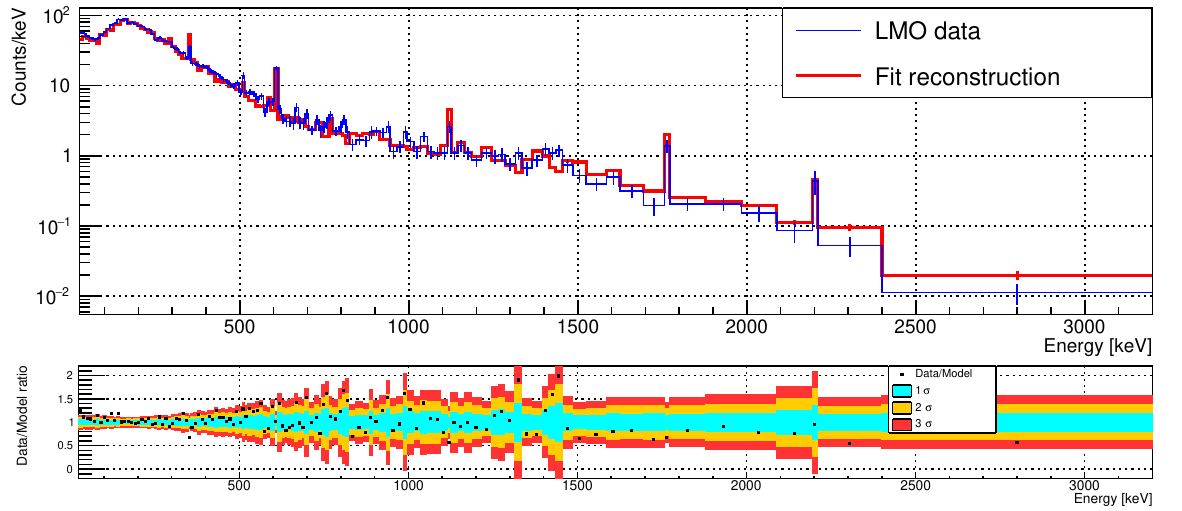}
\caption{Experimental data for CWO (top) and LMO (bottom) crystals compared with background model fit reconstruction. The lower panel shows the ratio between experimental counts and reconstruction counts for each bin. The colors indicates the uncertainty at $\pm1\sigma$, $\pm2\sigma$, and $\pm3\sigma$.}
\label{fig:bkg_fit}
\end{figure*}

Fig. \ref{fig:bkg_fit}  shows the result of the simultaneous fit of the CWO and the LMO experimental data to our background model, together with the ratio between the experimental number of counts and the model in each bin. All the resulting activities are compatible with the previous measured levels obtained by material screening\footnote{I. Bandac personal communication} and by assessing the $\alpha$ peaks in the case of the internal CWO contaminants. A small discrepancy is observed for $E < 500$ keV in the CWO spectrum and for $E<150$ keV in the LMO spectrum. 
In this latter, the lack of events in the fit reconstruction at low energies could be partially explained by the fact that we did not include any LMO internal contamination, in order to keep a reduced number of degrees of freedom. Our choice of not including the internal LMO radioactivities relies on the fact that we did not observe significant contaminations ($^{226}$Ra $<7$ $\mu$Bq/kg and $^{228}$Th  $<2$ $\mu$Bq/kg  \cite{Bandac:2023pkk}).

\begin{figure}[h!]
\centering
\includegraphics[width=0.45\textwidth]{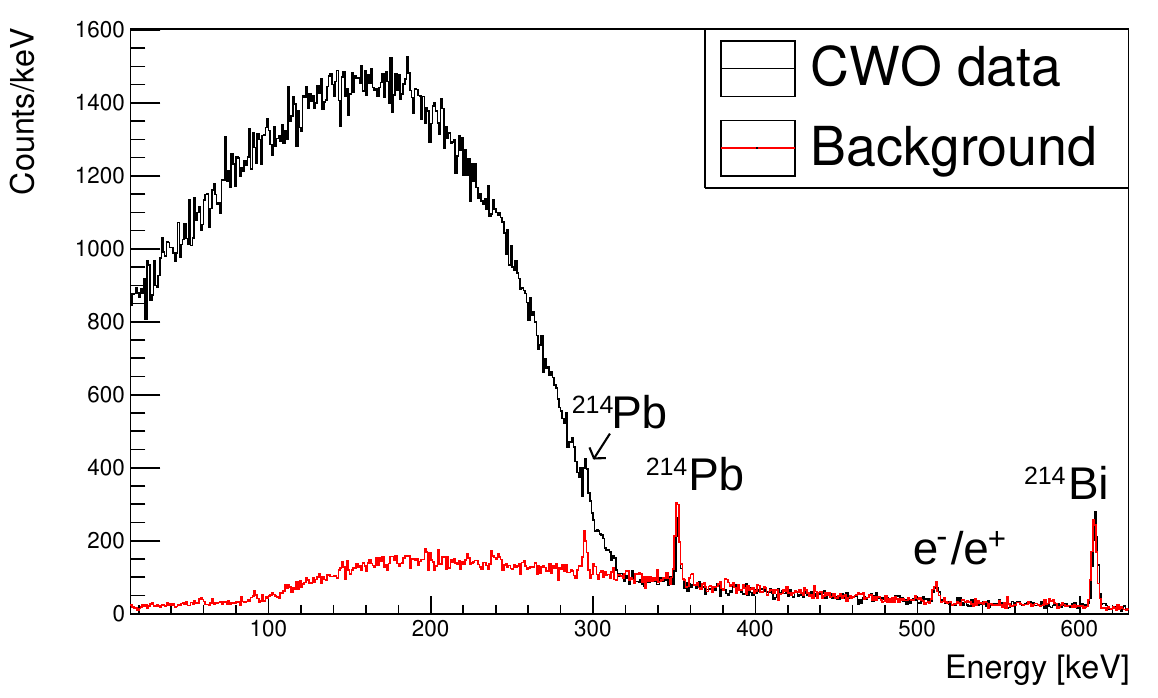}
\caption{Reconstruction of the experimental data from the background model fit in the region of the $^{113}$Cd $\beta$ decay compared to the experimental data. The signal-to-background ratio in the interval [15, 330] keV is ${\sim}12$.}
\label{fig:signal_to_bkg}
\end{figure}

We extrapolate the background spectrum in the $^{113}$Cd $\beta$ decay region, shown in Fig. \ref{fig:signal_to_bkg}, and found a signal-to-background ratio ${\sim}12$ in [15,~330] keV. In the following analysis, we account for a systematic uncertainty on the background spectrum to assess the uncertainty related to the simplicity of our chosen model. However, given the high signal-to-background ratio, this uncertainty has only a small impact on our results.

\section{Bayesian analysis}
\label{sec:Analyis}

The theoretical spectrum of the $^{113}$Cd $\beta$ decay was computed within the IBFM-2, MQPM and NSM models following the SSM decribed in Section \ref{sec:theory}. For each model, several spectra were computed for $g_A^{\text{eff}}$ values in the range [0.6,~1.4]  in step of 0.01, and s-NME in the range [$-$4.0,~3.0] with variable steps, where the smaller step is 0.05 in [$-$2.1,~$-$1.6] and [1.5,~2.0]. All spectra are computed with energy binning of 1 keV. 
The theoretical data were then convolved with the detector response, including the containment efficiency and the energy resolution. We performed a GEANT4 simulation of the detector response generating electrons with a uniform distribution in energy. The obtained spectrum is convolved with the theoretical spectra.  These effects are minor, given the high crystal density leading to a high containment efficiency and excellent energy resolution. 

We obtained the probability distribution function of the theoretical parameters, $\theta_i$, given the experimental data $\mathcal{D}$, from the Likelihood:
\begin{equation}
    \mathcal{L} = p(\mathcal{D}|\theta_i) = \prod_{i=1}^N \ \text{Pois}(n_i;\mu_i),
\end{equation}
where the product runs over bins $i$, $n_i$ is the observed number of counts in bin $i$, and $\mu_i$ is the reconstructed number of counts. The joint posterior distribution is sampled using the MCMC-based Bayesian Analysis Toolkit (BAT) \cite{Caldwell:2008fw}. 
An internal contamination of  $^{87}$Rb in the CWO crystal, from anthropogenic origin, was measured to $3$ mBq/kg based on an ICPMS measurement with an uncertainty at the level of ${\sim}30 \%$ \cite{Belli:2007zza}. Our background model can not constrain the contribution of this pure $\beta$ emitter (half-life = 5 $\times$ 10$^{10}$ yr, $Q_{\beta}$ = 282 keV), as its $Q$-value is lower than $323$ keV. To account for this contamination (or any low energy $\beta$ background that could affect the data), we consider this contribution as a separate parameter in our model: 

\begin{equation}
    \mu_i = s_i(E_i;g_A^{\text{eff}},\text{s-NME}) \cdot \epsilon(E_i) + B f_B(E_i) + R f_R(E_i),
\end{equation}
where $s_i$ is the convolved theoretical spectrum of $^{113}$Cd, $\epsilon$ is the analysis efficiency, $f_B$ is the background spectrum, $B$ is a normalisation factor of the background, and $f_R$ is the spectrum of the $^{87}$Rb $\beta$ decay with the normalisation factor $R$. At each step of the MCMC, we performed a 2D interpolation to extract the theoretical spectrum, $s_i$, for the given value of $g_A^{\text{eff}}$ and s-NME. 
We assigned a uniform prior between 0 and 100 mBq/kg to the $^{87}$Rb activity. The efficiency is fixed to the phenomenological function obtained in Fig. \ref{fig:efficiency}, and the parameter $B$ is allowed to float with a uniform prior. We perform the Bayesian fit in the interval [15,~1000] keV, with a binning of 1 keV. This interval allows to constrain the normalisation factor of the background spectrum. In the theoretical data, the s-NME can be either positive or negative. However, when allowing these two possibilities, the fit does not converge. Thus, we perform two separate fits for each model, one with positive s-NME and one with negative s-NME. 

\begin{figure*}[h!]
\centering
\includegraphics[width=1.0\textwidth]{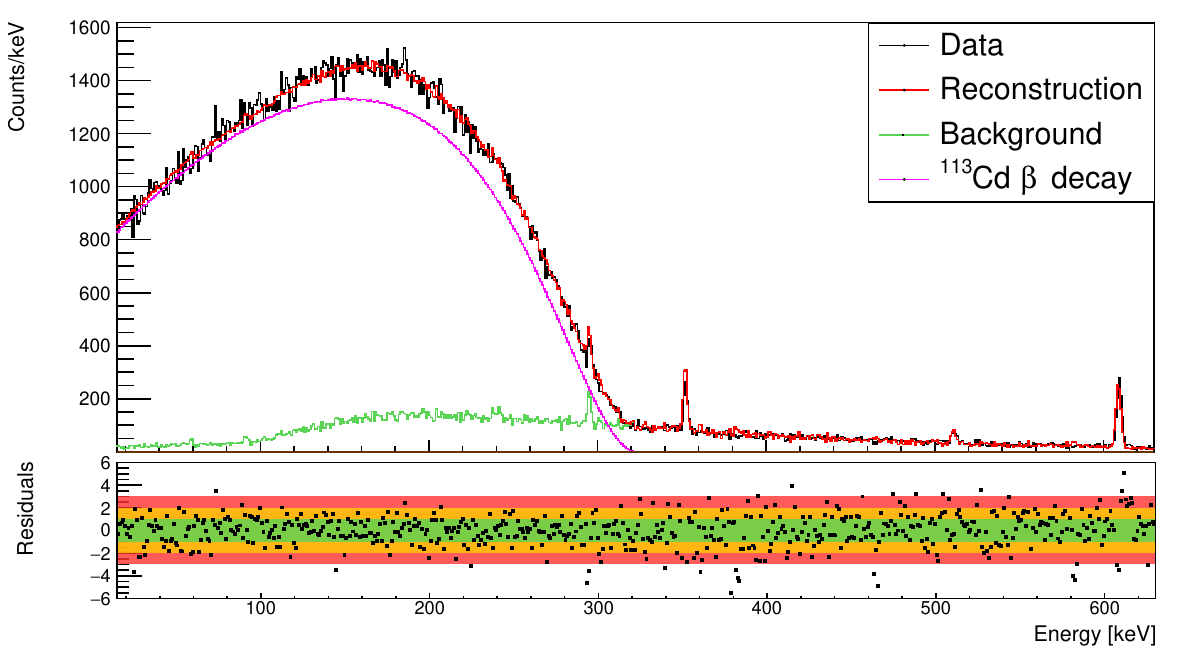}
\caption{Fit reconstruction of the $^{113}$Cd $\beta$ decay with the IBFM-2 model, for positive values of s-NME. The lower panel shows the residuals, which are defined as the difference between the number of counts in the data and the reconstruction divided by the square root of the number of counts in the data. The colors indicate the uncertainty at $\pm1\sigma$, $\pm2\sigma$, and $\pm3\sigma$. The experimental data are well reconstructed with $\chi^2 = 414$  for 305 degrees of freedom in the interval [15,~324] keV and no bias in the residuals. The reconstruction includes the $^{113}$Cd $\beta$ decay, the background model, and the crystal contamination of $^{87}$Rb, which activity has a marginalized posterior distribution compatible with 0 in this case, and thus, cannot be seen in the figure.}
\label{fig:CROSS_gA_reconstruction}
\end{figure*}

We show the fit reconstruction for the IBFM-2 model, with s-NME$>$0, in Fig. \ref{fig:CROSS_gA_reconstruction}. The data are well reconstructed with $\chi^2 = 414$ for 305 degrees of freedom, calculated in [15,~324] keV. The $\chi^2$ for the same number of degrees of freedom of the other models are reported in Table \ref{tab:gA_fit_results}. The p-value associated with each of the models is $\sim$0, which indicates that statistically the agreement is not perfect. This slight mismodelling of the data can come from the lack of statistics in some of the GEANT4 MC simulations that induces fluctuations in the background spectrum. These fluctuations do not affect the final measurement of the $^{113}$Cd $\beta$ decay spectral shape and the results concerning the $g_A^{\text{eff}}$, s-NME's and half-life values since we measure a continuum and we account for a systematic uncertainty related to the background model as presented below.


\begin{figure}[h!]
\centering
\includegraphics[width=0.47\textwidth]{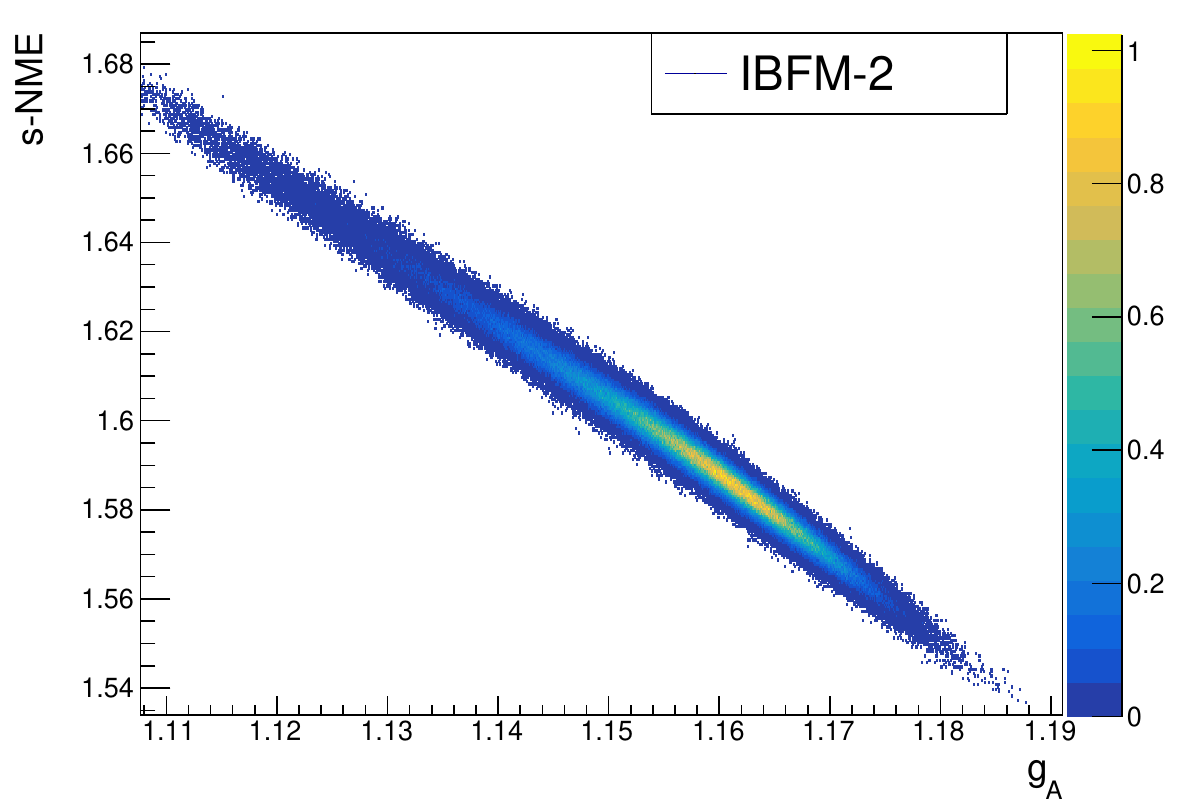}
\caption{Probability distribution function of the s-NME as function of $g_A^{\text{eff}}$  for the positive case of the IBFM-2 model. The color code shows the probability in arbitrary units. We observe a clear anti-correlation between the two parameters. }
\label{fig:CROSS_gA_NME_pdf}
\end{figure}

The two-dimensional posterior distribution of s-NME$>$0  as function of $g_A^{\text{eff}}$, for the IBFM-2 model is presented in Fig. \ref{fig:CROSS_gA_NME_pdf}, showing an anticorrelation between the two parameters.
From the fit we extract the $g_A^{\text{eff}}$ and s-NME values as the mode of the marginalized posterior distribution. We obtain the half-life by computing the integral of the theoretical spectrum at each step of the MCMC. 

 For each model and for each case  (positive s-NME and negative s-NME)  we considered systematic uncertainties to $g_A^{\text{eff}}$, s-NME's, and the half-life. For each systematic test, a probability distribution for the uncertainty is assumed based on the change in the best-fit value with respect to the reference fit. The total uncertainty is obtained by the convolution of these distributions with the statistical distribution of the reference fit. We consider the following systematic tests:
\begin{itemize}
    \item \textbf{Background sources localisation:} We perform tests to check the dependence of our results with the localization of the sources in the background model. We perform a fit considering only the most external sources (gap between the outermost cryostat screen and the external shield, environmental $\gamma$'s) and another fit considering only the closest sources (10 mK plate, copper holders). We include in both cases the contributions from the crystal, the lead shielding, and the pile-up. 
    We take a conservative approach choosing for the uncertainty the background model that gives the most important difference for each of the parameters ($g_A^{\text{eff}}$, s-NME and the half-life) and we assign a Gaussian posterior distribution to this systematic uncertainty. 
    \item \textbf{$^{87}$Rb:} It is observed that for the positive s-NME the $^{87}$Rb activity is compatible with 0 and gives a 90\% limit compatible with the activity that was reported in \cite{Belli:2007zza} ($3$ mBq/kg). For the negative case, the fit assigns a much larger activity, of the order of ${\sim}$50--80 mBq/kg. In the case of negative  s-NME, a contribution at low energies is needed to improve the fit (removing the $^{87}$Rb contribution we obtain $\chi^2$/d.o.f. = 1.9, 2.8 and 2.9 for the IBFM-2, MQPM and NSM model respectively). In \cite{Kostensalo:2023xzu}, it was observed that the positive s-NME reconstructs the experimental $^{113}$Cd $\beta$ spectrum better than the negative s-NME. Thus our data could indicate the same behavior.
    However, we can not rule out the presence of an additional pure $\beta$ decay contamination and therefore we keep the fits including the $^{87}$Rb contribution  as reference fits. To account for a systematic uncertainty we repeat the fit removing a contribution from $^{87}$Rb and we assign a uniform probability distribution to this uncertainty. 
    
    \item \textbf{Energy Bias:} The energy bias was corrected as mentioned in Section \ref{sec:measurement}. As a conservative choice, we performed fits varying the energy scale of the theoretical shapes by $\pm 1$ keV. The uncertainty is assigned with an asymmetric Gaussian distribution.
    \item \textbf{Bremsstrahlung:} To account for the uncertainty in the Bremsstrahlung cross-section, we performed GEANT4 simulations of the detector response modifying the cross-section by $\pm 10\%$ \cite{Pandola:2014uea}.
    These tests give exactly the same results as the reference fit with the precision we quote, thus the variance is well below the statistical fluctuations. Therefore this systematic can be neglected.
    \item \textbf{Efficiency:} Tests are done with an efficiency of $\pm 1 \sigma$ as shown in Fig. \ref{fig:efficiency}. The uncertainty is assigned with an asymmetric Gaussian distribution.
    \item \textbf{Energy resolution:} We consider systematic tests with an energy resolution of $\pm 1 \sigma$, as shown in Fig. \ref{fig:energy_resolution}. These tests give exactly the same results as the reference fit with the precision we quote, thus the variance is well below the statistical fluctuations. Therefore this systematic can be neglected.
\end{itemize}

The final results for the $g_A^{\text{eff}}$, s-NME, and the half-life, including the systematic effects, are reported in Table \ref{tab:gA_fit_results}.
\begin{table*}[h!]
\centering
\caption{Results of the Bayesian fit to the $g_A^{\text{eff}}$, s-NME and half-life of the $^{113}$Cd $\beta$ decay, including systematic uncertainties. The last column indicates the goodness of the fit expressed as $\chi^2$ with 305 degrees of freedom for the reference fit.}
\renewcommand{\arraystretch}{1.4}
\begin{tabular}{l|l|l|l|l}
Model                 & $g_A^{\text{eff}}$        & s-NME                    & $T_{1/2}$ [yr]       & $\chi^2$  \\
\hline
\hline
IBFM-2 (s-NME$> 0$)   & $1.160^{+0.019}_{-0.018}$ &  $1.59 \pm 0.04 $         & $7.50 \pm 0.08 $        & $414$ \\ 
IBFM-2 (s-NME$< 0$)   & $1.053^{+0.092}_{-0.088}$ &  $-1.74 \pm 0.13 $        & $7.84^{+0.28}_{-0.26} $ & $473$ \\ 
MQPM (s-NME$> 0$)     & $1.068 \pm 0.005$         &  $1.60 \pm 0.04 $         & $7.43 \pm 0.07 $        & $688$ \\ 
MQPM (s-NME$<0$)      & $1.050^{+0.055}_{-0.052}$ &  $-1.78^{+0.22}_{-0.21} $ & $8.10^{+0.40}_{-0.37} $ & $618$ \\ 
NSM (s-NME$>0$)     & $ 0.998 \pm 0.006 $       &  $1.66 \pm 0.03 $         & $7.45 \pm 0.08 $        & $572$ \\ 
NSM (s-NME$< 0$)    & $0.987^{+0.070}_{-0.068}$ &  $-1.86^{+0.22}_{-0.20} $ & $8.07 \pm 0.38 $        & $608$\\ 
\hline          
\end{tabular}
\label{tab:gA_fit_results}
\end{table*}
All results indicate a small renormalisation of the $g_A^{\text{eff}}$ with values between 1.0 and 1.2. The s-NME are between 1.6 and 1.7 for the positive case and $-$1.9 and $-$1.7 for the negative case. The values we obtained here with the spectrum-shape method show an extremely good agreement with the ones relying on a different approach, the spectral moments method (SMM) presented in \cite{Kostensalo:2023xzu}. 
In the SMM, the fitting of spectral moment $\mu_0$ that corresponds to the decay rates and $\mu_1$ that corresponds to the mean energy of the spectrum allows to constrain the $g_A^{\text{eff}}$ and s-NME. The experimental data in \cite{Kostensalo:2023xzu}  are based on \cite{Belli:2019bqp}. We can also compare our results to the ones obtained by the COBRA experiment \cite{Kostensalo:2020gha}, that used the same theoretical framework but another type of detector technology (CdZnTe semiconductor detector) and analysis. In \cite{Kostensalo:2020gha}, the analysis was done in two steps. The first step consists of finding the combination of $g_A^{\text{eff}}$ and s-NME that match the experimental half-life measured in \cite{Belli:2007zza}. Then, a spectral shape comparison was performed based on a $\chi^2$ test between their experimental data and the theoretical spectra. This second step was giving another set of combinations of values of $g_A^{\text{eff}}$ and s-NME. The combination that was reconstructing the half-life, as well as the spectral shape, was finally extracted. Our results on the $g_A^{\text{eff}}$ are compatible with the value obtained by COBRA within 1.4$\sigma$ for NSM, 1.2$\sigma$ for MQPM, and  2.3$\sigma$ for IBFM-2, where our results are systematically higher than the COBRA values.


In our analysis, as the three models must reconstruct the half-life, we combined the results by computing the mean of $T_{1/2}$, while the uncertainty is obtained by randomly choosing an uncertainty value in each distribution and summing them, leading to a final value of:
\begin{equation}
    T_{1/2} = 7.73 ^{+0.60}_{-0.57} \times 10^{15} \ \text{yr}.
    \label{eq:CWO_demi_vie_final}
\end{equation}
This is a conservative choice as we sum the uncertainties of the different nuclear models that are correlated. This value is well-compatible with the previous measurements \cite{Danevich:1996,Goessling:2005,Belli:2007zza,Dawson:2009ni}. The half-life corresponds to a number of counts of $5.75 ^{+0.45} _{–0.42} \times 10^{5} $, in \cite{Belli:2007zza} the total number of counts was $(24.03 \pm 0.01) \times 10^{5}$ .


We also fitted our $^{113}$Cd $\beta$ spectral shape with a phenomenological function, as described in \cite{Belli:2007zza}. The aim of this fit is to improve event generators providing beta spectral shapes, like Decay0 \cite{Ponkratenko:2000um}, which use phenomenological functions to generate initial kinematics, instead of theoretical calculations. It may also be useful to compare future models to our data by means of a phenomenological function including the systematic uncertainties. From the original CWO spectrum, we subtracted the background model obtained in Section \ref{sec:BM} and corrected by the event selection efficiency shown in Fig. \ref{fig:efficiency}. We then fitted the spectral shape with the Eq. \ref{eq:general_beta_decay}, with an arbitrary normalization, and with the phenomenological shape factor used in \cite{ALESSANDRELLO1994394, Belli:2007zza}:
\begin{equation}
    C(w_e) = p_e^6 + 7 a_1 p_e^4 q^2 + 7 a_2 p_e^2 q^4 + a_3 q^6,
\end{equation}
where $p_e$ and $q$ are the momentum of the emitted electron and neutrino, respectively. This expression is characteristic of three-fold forbidden unique $\beta$ decay \cite{konopinski1966theory, Mougeot:2015bva}. However, it was found to be able to describe the $^{113}$Cd $\beta$ spectral shape \cite{Belli:2007zza, ALESSANDRELLO1994394} and it can be used to reconstruct our observed $^{113}$Cd $\beta$ decay spectral shape. We applied the energy resolution directly in the fit function and considered the $Q_{\beta}$, noted $w_0$ in Eq. \ref{eq:general_beta_decay}, as a free parameter. The data are well reconstructed with $\chi^2 = 400$ for 304 degrees of freedom. We evaluated the same systematic uncertainties except for the one related to the $^{87}$Rb and obtained:
\begin{align*} 
    &Q_{\beta} = (322.8 \pm 1.1) \text{ keV}, \\
    &a_1 = 1.08 \pm 0.03, \\
    &a_2 = 1.69 \pm 0.04, \\
    &a_3 = 4.11 \pm 0.15, \\
    &T_{1/2} = (7.56 \pm 0.08) \times 10^{15} \text{ yr}.
\end{align*}
The $Q_{\beta}$ is in excellent agreement with the accepted AME2020 value of $Q_{\beta} = 323.84(27)$ keV \cite{Wang:2021xhn}. We can compare these results with the following $a_i$ coefficients obtained in \cite{Belli:2007zza}:
\begin{align*} 
    &a_1 = 1.016 \pm 0.005, \\
    &a_2 = 1.499 \pm 0.016, \\
    &a_3 = 3.034 \pm 0.045. \\
\end{align*}
The $a_3$ coefficient disagrees with our value, however this parameter is highly sensitive to the energy threshold, which is 15 keV in our analysis, while it was 28 keV in \cite{Belli:2007zza}, and can explain the difference.

\section{Conclusion}

We performed a precise spectral shape measurement of the $^{113}$Cd $\beta$ decay using a 0.43 kg CdWO$_4$ crystal operated as a bolometer in low background conditions of the CROSS facility at the Canfranc underground laboratory in Spain. We extracted the spectrum of the CdWO$_4$ crystal using 634 hours of data, and we measured the energy resolution, the energy bias, and the events selection efficiency. We then constructed a background model of the CdWO$_4$ detector, further constrained thanks to a $^{100}$Mo-depleted Li$_2$MoO$_4$ scintillating bolometer operated in the same set-up. The background obtained in this way in the CdWO$_4$ detector was extrapolated to the $^{113}$Cd region of interest [15,~330] keV and used as an input in a spectral shape fit based on a Bayesian method. 
The $^{113}$Cd decay spectral shape provides a sensitive test of nuclear models in a regime of large angular momentum differences, similar to $0\nu\beta\beta$ decay which has a large momentum transfer and proceeds via higher-multipolarity intermediate nuclear states. We used theoretical spectra based on the spectrum-shape method for three nuclear models, IBFM-2, MQPM and NSM. Our fit has two free parameters, the effective weak axial-vector coupling constant, $g_A^{\text{eff}}$, and the small relativistic nuclear matrix elements, s-NMEs. 
Our fits suggest that the positive s-NMEs reconstruct the experimental data better than the negative s-NMEs. We derived  best fit values of the parameters $g_A^{\text{eff}}$ and s-NME, and obtained  $g_A^{\text{eff}}$ between 1.0 and 1.2, indicating possibly a deficiency of the nuclear models. 
Our best fit values of $g_A^{\text{eff}}$ and s-NMEs are in excellent agreement with the values obtained by \cite{Kostensalo:2023xzu} within the spectral moments method, which relies on a different approach and used experimental data based on \cite{Belli:2019bqp}.  Our results on the $g_A^{\text{eff}}$ are compatible but systematically higher than the results from the COBRA experiment \cite{Kostensalo:2020gha}, that used the same theoretical framework. They are within 1.4$\sigma$ for NSM, 1.2$\sigma$ for MQPM, and  2.3$\sigma$ for IBFM-2.
We measured the $^{113}$Cd half-life as $T_{1/2} = 7.73 ^{+0.60}_{-0.57} \times 10^{15} \ \text{yr}$, including systematic uncertainties, compatible with the previous measurements \cite{Danevich:1996,Goessling:2005,Belli:2007zza,Dawson:2009ni}. 
We could also describe the $^{113}$Cd $\beta$ decay spectral shape with a phenomenological equation of three-fold forbidden unique $\beta$ decay, reconstructing with good precision the $Q_{\beta}$.

\section{Acknowledgments}
This work is supported by the European Commission (Project CROSS, Grant No. ERC-2016-ADG, ID 742345) and by the Agence Nationale de la Recherche (Project CUPID-1; ANR-21-CE31-0014, ANR France). We acknowledge also the support of the P2IO LabEx (ANR-10-LABX0038) in the framework ”Investissements d’Avenir” (ANR-11-IDEX-0003-01 – Project ”BSM-nu”) managed by ANR, France. The INR NASU group was supported in part by the National Research Foundation of Ukraine Grant No. 2023.03/0213.

\bibliographystyle{spphys}       
\bibliography{biblio}

\end{document}